\documentclass[aps,pre,reprint,twocolumn,showpacs,floatfix,nofootinbib,
 superscriptaddress,
 ]{revtex4-1}
\usepackage{subfigure}
\usepackage{amssymb}
\usepackage{amsfonts}
\usepackage{amsmath}
\usepackage{amsthm}
\usepackage{epsfig}
\usepackage{graphicx}
\usepackage[usenames,dvipsnames]{color}
\usepackage[latin1]{inputenc}

\usepackage[hidelinks,unicode=true]{hyperref}
\hypersetup{colorlinks=true,
	linkcolor=blue,
	urlcolor=blue,
	citecolor=blue,
	pdfhighlight=/N
}

\usepackage{comment}
\usepackage[normalem]{ulem}

\usepackage[dvipsnames]{xcolor}

\newcommand{\av}[1]{\langle{#1}\rangle}
\renewcommand{\vec}{\mathbf}
\newcommand{\vinf}{v_\infty}

\begin{document}
\title{Eden model with nonlocal growth rules and the kinetic roughening in biological systems}
\author{Silvia N. Santalla}
\affiliation{Departamento de F\'{\i}sica and Grupo Interdisciplinar de Sistemas
  Complejos, Universidad Carlos III de Madrid, Legan\'es, Spain}
\author{Silvio C. Ferreira}
\affiliation{Departamento de F\'{\i}sica, Universidade Federal de Vi\c{c}osa,
36570-900, Vi\c{c}osa, Minas Gerais, Brazil}
\affiliation{National Institute of Science and Technology for Complex Systems, Rio de Janeiro, Brazil}

\begin{abstract} 
We investigate an off-lattice  Eden model where the growth of new cells is
performed with a probability dependent on the availability of resources coming
externally towards the growing aggregate. Concentration of nutrients necessary
for replication is assumed to be proportional to the voids connecting the
replicating cells to the outer region,  introducing therefore a nonlocal
dependence on the replication rule. Our simulations point out  that the
Kadar-Parisi-Zhang (KPZ) universality class is a transient  that can last for
long periods in plentiful environments. For conditions of nutrient scarcity, we
observe a crossover from regular KPZ to unstable growth, passing by a transient
consistent with the quenched KPZ class at the pinning transition.  Our analysis
sheds light on results reporting on the universality class of kinetic roughening
in akin experiments of biological growth.
\end{abstract}



\maketitle

\section{Introduction}
\label{sec:intro}
Pattern formation in biological growth is a nonequilibrium process that has
attracted a lot of attention~\cite{reichl1998,Meakin,Murray2}. Particular interest is given to the rough interfaces of compact and spherical patterns  observed
in bacterial colonies~\cite{Fujikawa89,Vicsek90,Wakita1997,Matsushita98,santalla17} and
clusters of normal~\cite{Huergo10,Huergo11,Huergo14} and
tumor~\cite{Bru98,Huergo12} cells grown on culture under controlled experimental
conditions.

The Eden model~\cite{Eden61} is a benchmark  of stochastic processes, in the
important class of growth models on expanding
substrates~\cite{Pastor2007,Escudero2008,Escudero2009,Escudero2011,Masoudi2012,Escudero2013,Carrasco2014}, which forms radial clusters with irregular (fractal) borders. In this model, new cells are irreversibly added at random positions of the neighborhood of previously existent cells. Off-lattice simulations of the radial growth of the Eden model~\cite{Ferreira06,Alves11,Takeuchi12,Alves2012,Alves13} show that it belongs to the Kardar-Parisi-Zhang (KPZ)~\cite{KPZ} universality class introduced with the equation
\begin{equation}
\frac{\partial h}{\partial t} = F+\nu \nabla^2h +\frac{\lambda}{2}|\nabla h|^2+\eta(\vec{x},t)
\label{eq:KPZ}
\end{equation}
in studies of nonequilibrium kinetic roughening. In Eq.~\eqref{eq:KPZ}, $h$ is
the distance to the initial $d$-dimensional flat substrate, $\eta$ is a 
uncorrelated Gaussian noise given by
$\av{\eta(\vec{x},t)\eta(\vec{x}',t')}=2D\delta^d(\vec{x}-\vec{x}')
\delta(t-t')$, and $F$ is a constant flux driven the surface front. The
Laplacian accounts for surface tension while the quadratic term produces an
excess velocity and represents lateral growth along the normal direction in the
substrate, a  hallmark of the KPZ universality class. For radial growth the
height is replaced by the distance $r$ to the origin, which is the place where
the process usually starts. The Eden model is a Markovian, local growth process
just as the KPZ equation. If the white noise of Eq.~\eqref{eq:KPZ} is replaced
by a position dependent noise ${\eta}_q(\mathbf{x},h)$ we obtain the quenched
KPZ (qKPZ) equation~\cite{Parisi1992,Barabasi95}
\begin{equation}
\frac{\partial h}{\partial t} = F+\nu \nabla^2h +\frac{\lambda}{2}|\nabla h|^2+\eta_q(\vec{x},h).
\label{eq:qKPZ}
\end{equation}
In the qKPZ equation, the surface can be
pinned if the driving force is below a threshold $F_c$~\cite{Barabasi95}.

The basic characterization of self-affine kinetic roughening can be done by 
means of the growth, roughness, and dynamical exponents $\beta$, $\alpha$, and
$z$, respectively.  The first one is related to the evolution of the variance of the
surface radii (or heights in a flat initial condition)  given by
\begin{equation}
W^2= \av{r^2}_c = \av{r^2}-\av{r}^2\sim t^{2\beta},
\label{eq:rugo}
\end{equation}
where the notation $\av{r^n}_c$ represents the $n$th cumulant of $r$ averaged
over both the whole surface and independent samples. In the absence of anomalous
scaling~\cite{Lopez1997a}, the second exponent gives the point-to-point
fluctuations along the surface and can be calculated from the local interface
width within sectors\footnote{For a flat substrate of constant size $L$, the
	roughness exponent can be defined in terms of asymptotic value of the interface
	width that saturates at long times as $W^2_\text{sat}\sim L
	^{2\alpha}$~\cite{Barabasi95}. However, in the inflating geometry investigated
	in the present work, the interface width does not saturate and this definition
	loses meaning. See also discussions in Refs.~\cite{Pastor2007,Carrasco2014}.} of
length $l$ as $w_\text{loc}^2\sim l^{2\alpha}$~\cite{Barabasi95}. Finally, the
third exponent gives the spreading of the characteristic surface correlation
length as $\xi\sim t^{1/z}$. In case of self-affine growth the exponents obey
the scaling relation $\alpha=\beta z$~\cite{Barabasi95}. The KPZ
	equation has exponents $\beta=1/3$, $\alpha=1/2$ and $z=3/2$ in
	$d=1$~\cite{KPZ}, which are observed in numerical simulations of several
	models~\cite{Meakin,Barabasi95}. These exponents also hold for qKPZ equation
above the critical driving force, $F>F_c$. Exactly at the pinning transition,
the exponents are given by $\alpha\approx \beta \approx
0.633$~\cite{Barabasi95,Odor2008} which are related to the directed percolation
universality class~\cite{Tang95}.

A breakthrough on the field was surveyed by the exact one-point solutions of the
one-dimensional KPZ equation~\cite{Sasamoto10,Sasamoto10b,Amir11,Calabrese11}.
It was shown that a surface point asymptotically evolves as\footnote{The ansatz
	of Eq.~\eqref{eq:ansatz} was firstly considered in Ref.~\cite{Krug92} and obtained later
	in analytical solutions of the polynuclear growth model~\cite{Prahofer00} and single
	step model~\cite{Johansson00}. Chronological events related to progress of the KPZ
	class can be found elsewhere~\cite{Halpin15}.}
\begin{equation}
r\simeq v_\infty t + s_\lambda (\Gamma t)^\beta \chi,
\label{eq:ansatz}
\end{equation}
where $s_\lambda =\text{sgn}(\lambda)$, $v_\infty$ is the asymptotic
average velocity of the surface, $\Gamma$ is a system dependent 
parameter and $\chi$ is a stochastic variable that follows the
Tracy-Widom (TW) probability distribution function~\cite{Tracy94} of
the largest eigenvalue of the Gaussian unitary ensemble (GUE) for radial
growth (or Gaussian orthogonal ensemble for flat  geometries).

Contrasting with the plenty of models reporting great accordance with the KPZ
exponents in $d=1$~\cite{Meakin,Barabasi95}, the  experimental
	counterparts were just a few~\cite{Maunuksela1997,Schilardi1999,Takeuchi2013}.
The situation has changed recently after a thorough statistical analysis of
patterns produced between distinct turbulent phases of liquid
crystals~\cite{Takeuchi10,Takeuchi2011,Takeuchi2012} which was followed by other
experiments reporting KPZ
exponents~\cite{Yunker2013,Atis2013,Atis2015,Huergo10,Huergo11,Huergo12}.
Regarding biological growth, essentially all experimental essays on compact
bacteria colonies~\cite{Fujikawa89,Vicsek90,Matsushita98,santalla17} or cell
aggregates~\cite{Bru98,Bru03} ruled out KPZ universality class until quite
recently when \textit{in vitro} experiments with eukaryotic cells, namely
\textit{Vero}~\cite{Huergo10,Huergo11} and cancer HeLa~\cite{Huergo12} cells,
yielded exponents in pretty good agreement with the KPZ class. The necessary
conditions to observe the KPZ exponents are high renewal rate of the culture
medium granting sufficient nutrients for cells and removing the waste products
produced by the cellular processes and soft medium permitting cell motility.
When the same cell colonies are grown on highly viscous gelled media, the
exponents change to $\beta\approx 0.75$ and $\alpha\approx 0.63$, which were
associated with the qKPZ class at the pinning transition~\cite{Huergo14}.

A fundamental difference between the Eden model and biological
growth is the locality inherent to the former, which is generally absent in the
latter. The cell replication and motility depend on several factors including
nutrient resources and intercellular signaling mediated by
proteins~\cite{Freshney10}. The chemicals diffuse and react (are absorbed or
released by cells) in a culture medium such that cellular responses are
diffusion-limited and, in principle, nonlocal processes. So, the experimental
evidences for KPZ presented by Huergo \textit{et al}.~\cite{Huergo10,Huergo11,Huergo12} are, in principle, surprising and one
would wonder how sensitive to the experimental conditions is this kind of
experiments.

In this paper, we tackle this problem investigating an off-lattice Eden model
with a nonlocal growth rule, in which the replication of a cell depends on the
empty space bridging its position at the colony and the outer medium of the
culture. This simple shadowing model mimics inward nutrient diffusion and was introduced
in Ref.~\cite{santalla17} for the modeling of bacteria colonies. We observe
that the KPZ regime can be seen for long times if the shadowing effects are
weak. Under resource scarcity a crossover from regular KPZ to unstable growth
passing by a transient consistent with qKPZ class at the pinning transition can be
observed. Therefore our results help to understand the recent observations of
KPZ class on biological growth~\cite{Huergo10,Huergo11,Huergo12,Huergo14}.

The sequence of the paper is organized as follows. The model and its computer
implementation are presented in section~\ref{sec:model} while the results
obtained from numerical simulations are presented in section~\ref{sec:results}.
Our conclusions and prospects are drawn in section~\ref{sec:conclu}.

\section{Model and methods}
\label{sec:model}

\begin{figure}[hbt]
	\centering
	\includegraphics[width=0.7\linewidth]{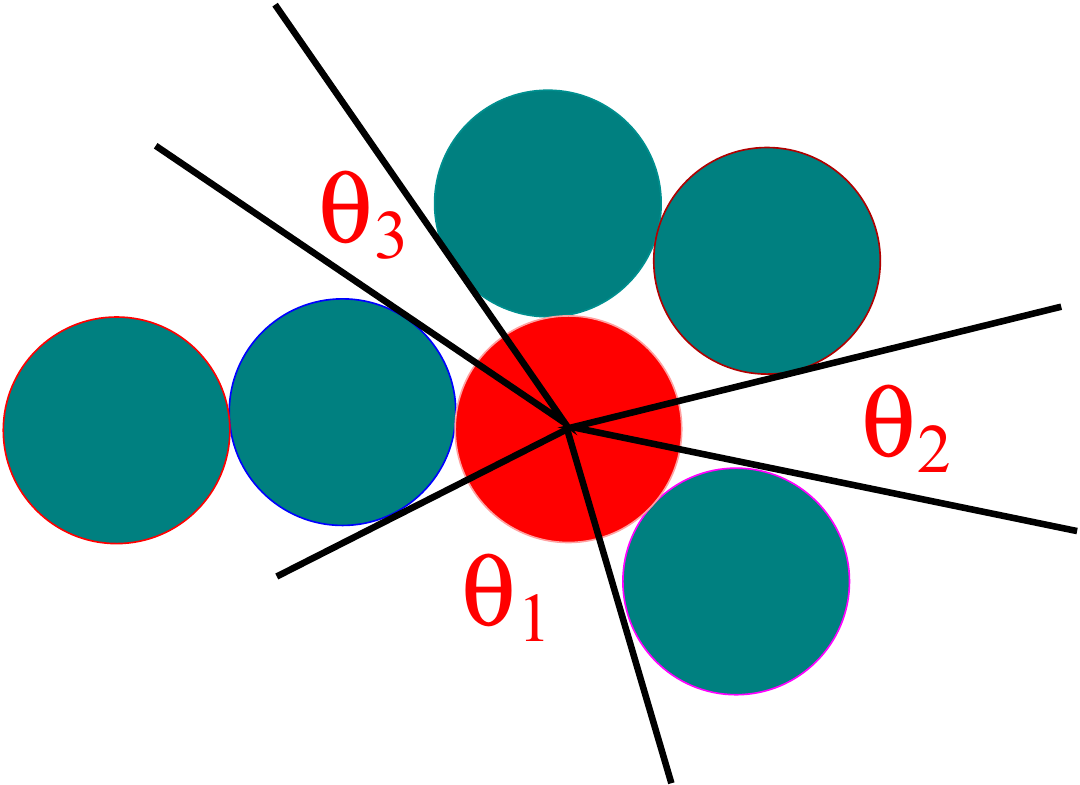}
	\caption{Determination of the aperture angle $\Theta_s$ ruling the replication
		probability of the model and given by Eq.~\eqref{eq:shadow}. The red cell is the
		replicating one. The angle is given by the sum of all available
		apertures, $\Theta_s=\sum_i \theta_i$.}
	\label{fig:model}
\end{figure}

We perform simulations in two dimensions of a modification of the off-lattice
Eden model introduced in Ref.~\cite{Takeuchi12}. The cells are represented by
discs of diameter $a=1$. The simulation starts with a single cell at the origin.
New cells are added one at a time according to the following rules. A cell of
the aggregate is chosen at random with equal chance and accepted with
probability $P$ (originally we have $P=1$~\cite{Takeuchi12,Alves13}). A
randomly selected position, tangent to the parent cell, is chosen and a daughter
cell is added at this position if this attempt does not produce an overlap
with any other cell. The time is incremented by $\Delta t=1/N_\text{cell}$,
where $N_\text{cell}$ is the number of cells in the aggregate, independently if
a new cell was added or not.

\begin{figure}[hbt]
	\centering
			\boxed{\includegraphics[width=0.45\linewidth]{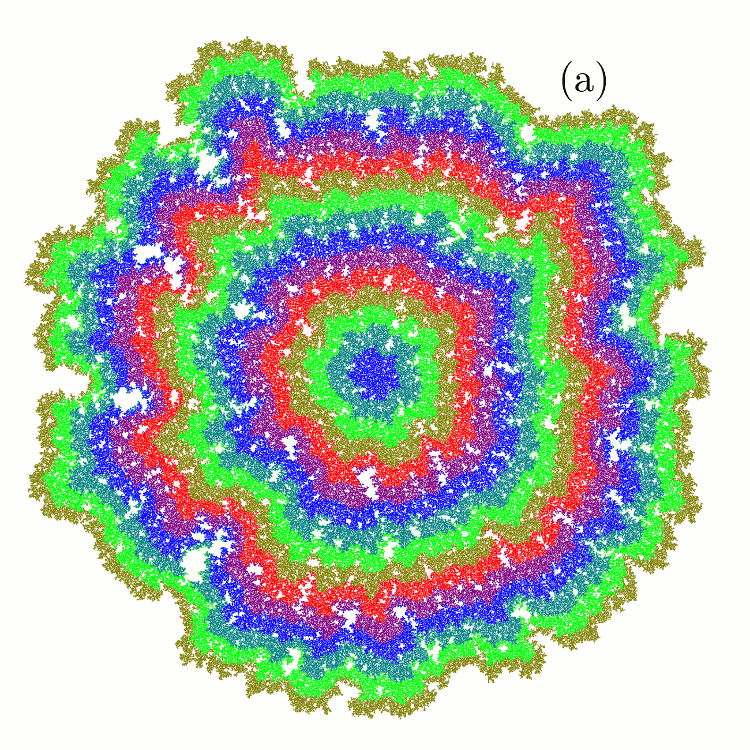}}\\~\\
			\boxed{\includegraphics[width=0.43\linewidth]{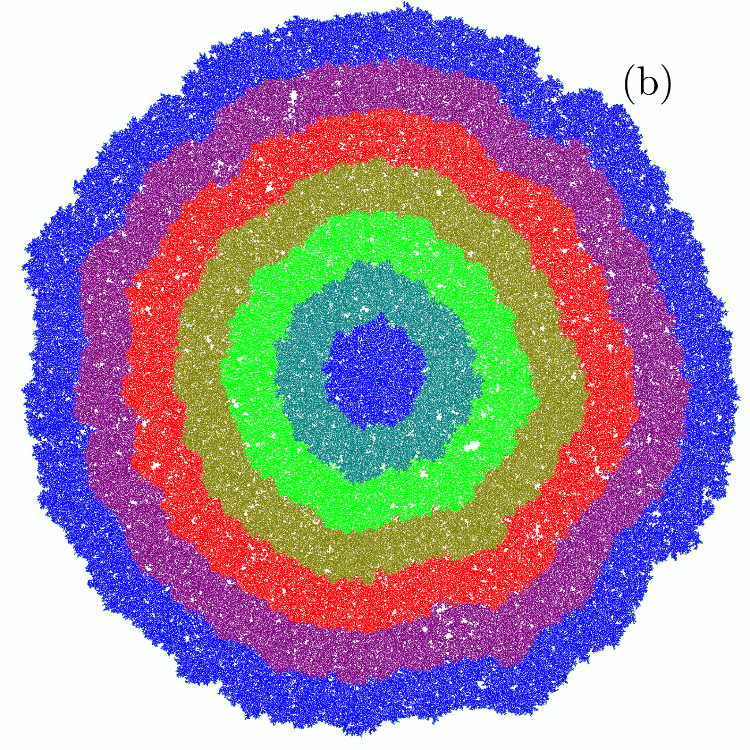}}~~
	       	\boxed{\includegraphics[width=0.43\linewidth]{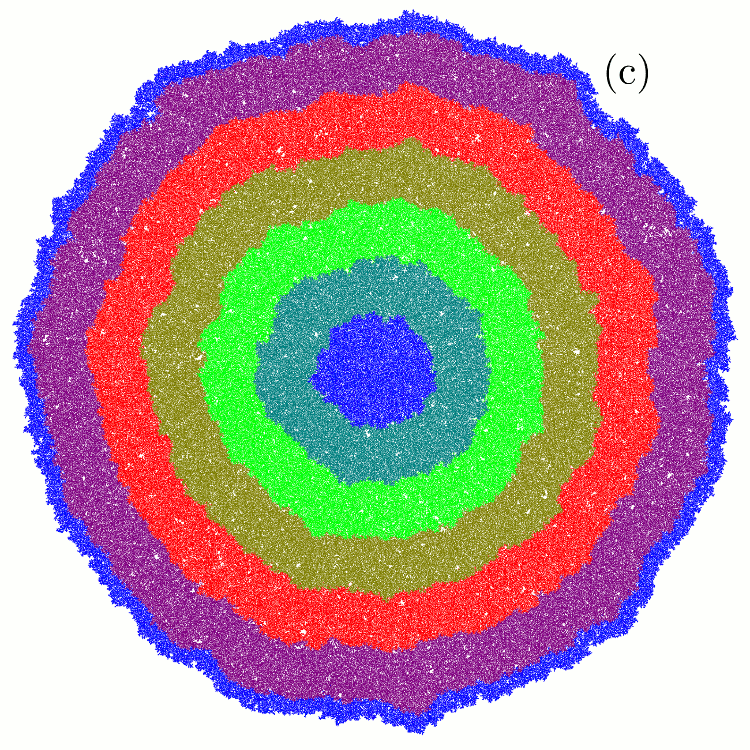}}
	\caption{Aggregates for different levels of shadowing: (a) $A_s=1$ (strong),
		(b) $A_s=5$ (intermediary) and (c) $A_s=9$ (weak). Colors change every 150 time
		unities. Square regions correspond to a length $1000a$. Optimizations were not used in these simulations.}
	\label{fig:agregados}
\end{figure}

The non-local effects are included with the probability
\begin{equation}
 P=1-\exp\left(-A_s {\Theta_s\over 2\pi}\right),
\label{eq:shadow}
 \end{equation}
where $\Theta_s$ is the total local aperture angle defined as the fraction of
rays emanating from the center of the progenitor cell that do not intersect the
aggregate as illustrated in Fig.~\ref{fig:model}. The parameter  $A_s$ controls
the  shadowing effect strength. This probability simulates in a simple fashion
the dependence on the nutrients that are scarcer at the innermost regions of the
aggregate. The higher the value of $A_s$ the higher the density of nutrients
implying that a small value of $A_s$ would require a large free region to permit
replication.

We consider two types of simulations. In the first one, we run 2000 samples and
clusters with diameter $10^3a$.  In the second one, 200 samples and a diameter
of $5 \times 10^3a$ are used. Simulations in the second group were optimized
removing from the list of cells elective for replication those that did not
replicate after $M=40$ attempts. We varied this number of attempts and
observed that simulations become independent of this parameter for $M\ge40$.  In
both cases, the outer borders are converted into single-valued profiles with
respect to the angle, permitting the analysis of self-affine surfaces.

Figure \ref{fig:agregados} shows three aggregates for different values of $A_s$.
Spherical aggregates with rough borders, resembling those of the original Eden
model, are obtained for large values of $A_s$ or, equivalently, a weak shadowing
effect. For small $A_s$,  where shadowing is strong,  the surface presents more
overhangs forming protrusions that resemble unstable growth of reaction-diffusion
models with explicit modeling of nutrient concentration fields by means of
continuous diffusion equations~\cite{Goldin98,Lacasta99,Kozlovsky99,Ferreira02}.

\section{Results}
\label{sec:results}

\begin{figure}[ht]
	\centering
	\includegraphics[width=0.75\linewidth]{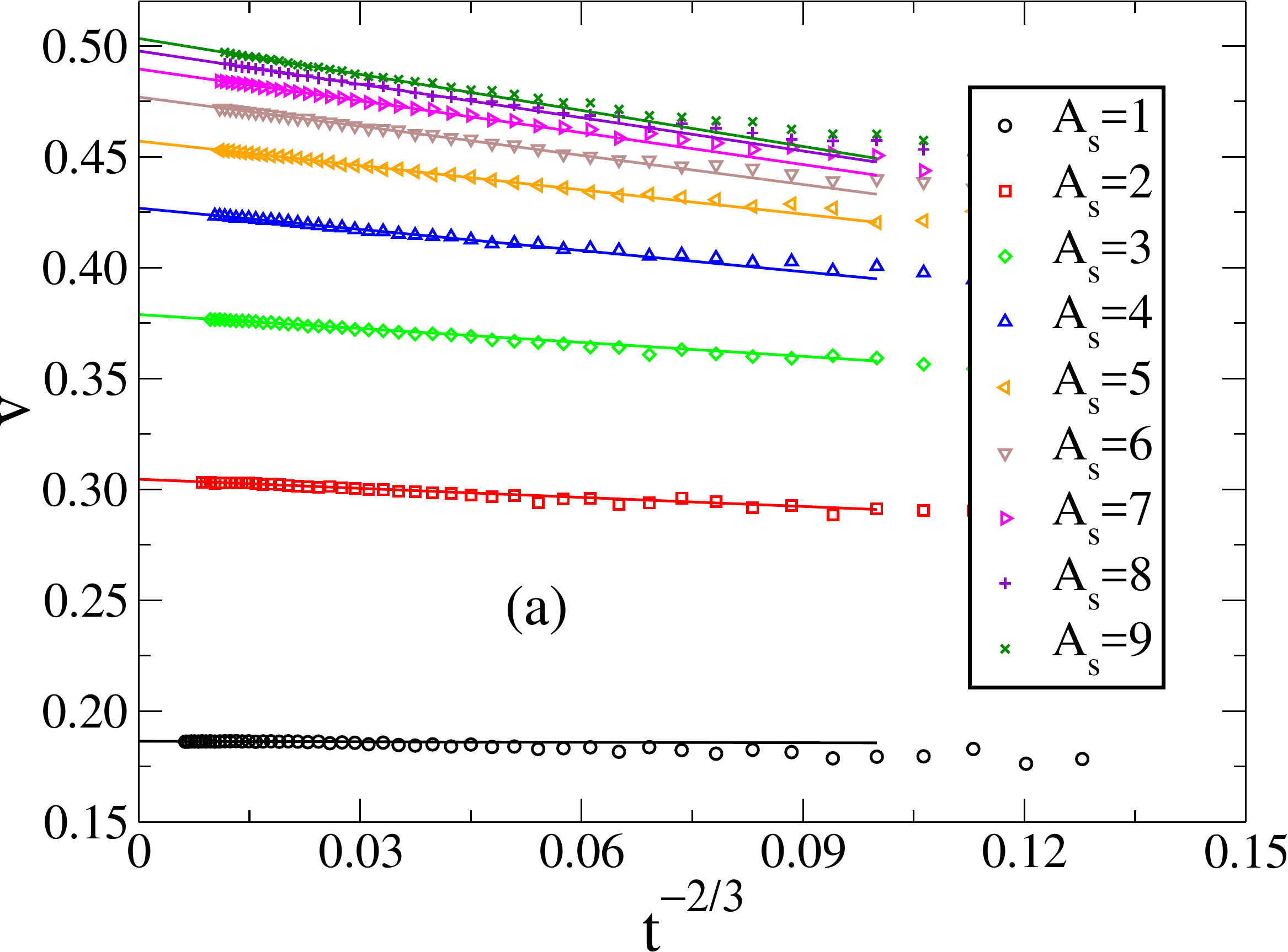}\\~\\
	\includegraphics[width=0.78\linewidth]{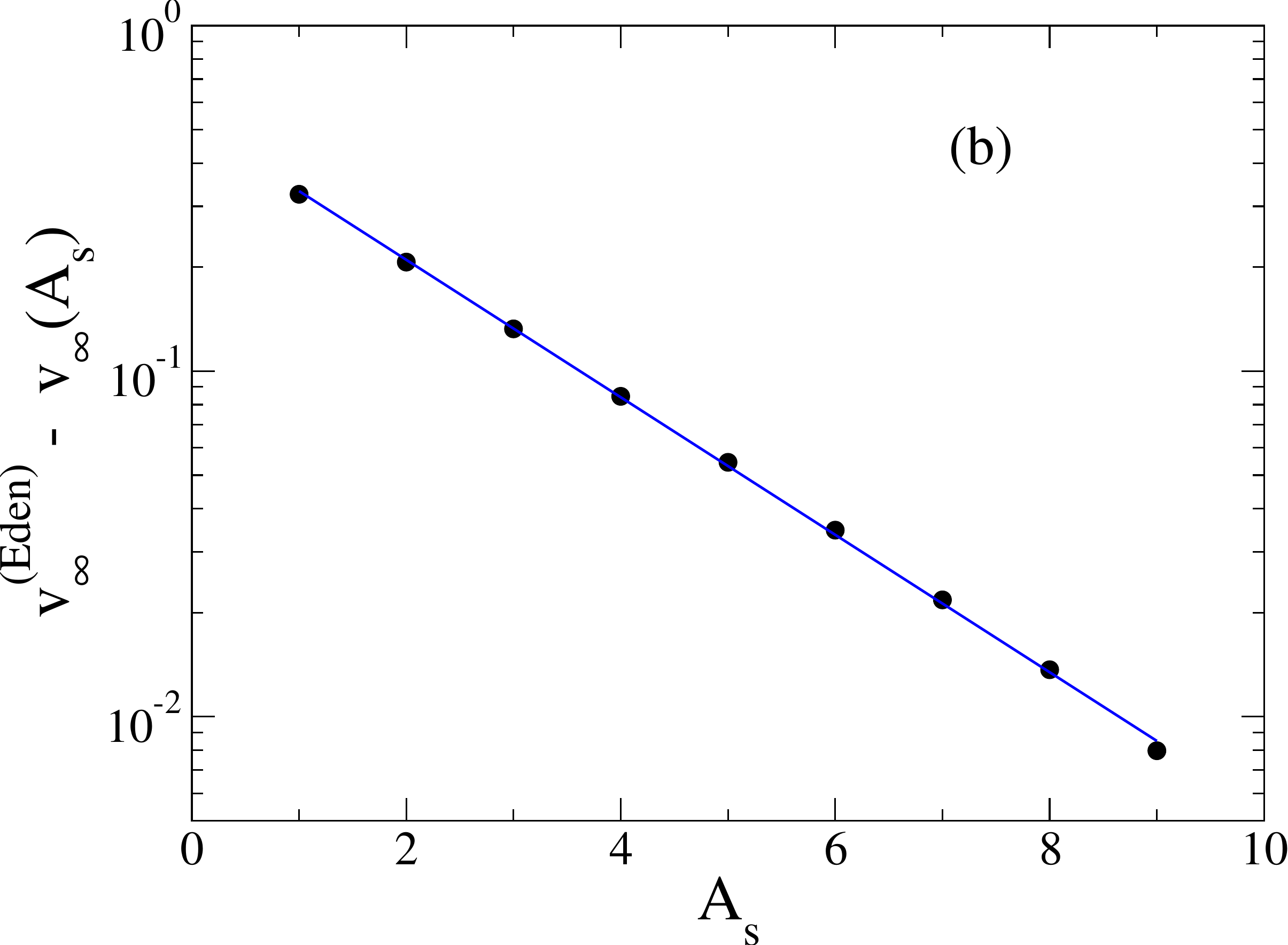}
	\caption{(a) Surface velocity against $t^{-2/3}$ for different values of the
		parameter $A_s$. Lines are linear regressions. (b) Asymptotic interface velocity
		for the modified Eden model as a function of the shadowing parameter. Line is an
		exponential regression. Optimizations were not used in these simulations. }
	\label{fig:velocity}
\end{figure}

The interface velocity in the KPZ regime given by \eqref{eq:KPZ} reads as
 \begin{equation}
 v=\frac{d\av{r}}{dt}=\vinf+\beta \Gamma^\beta t^{-1+\beta},
 \end{equation}
so that velocity plotted against the $t^{-1+\beta}$ should be a linear curve.
These plots are shown in Fig.~\ref{fig:velocity} using
$\beta=\beta_\text{KPZ}=1/3$ for all investigated parameters. We can see the 
linear behavior expected in the KPZ regime  for week shadowing (large $A_s$)
which means a correction $t^{-2/3}$ in the velocity. For strong shadowing,
corrections in the asymptotic velocity are negligible, being apparently null and
corresponding to a non KPZ regime. For the investigated range of the parameter
$A_s$, the asymptotic velocity depends on this parameter as
 \begin{equation}
 \vinf \simeq \vinf^\text{(Eden)} - 0.525 \exp({-0.458A_s}),
 \end{equation}
where $\vinf^\text{(Eden)}=0.51371$ is the asymptotic velocity of the pure Eden
model~\cite{Alves13}. This exponential dependence of the asymptotic velocity
with the shadowing parameter $A_s$ can be seen in Fig.~\ref{fig:velocity}(b).

\begin{figure}[ht]
	\centering \includegraphics[width=0.7\linewidth]{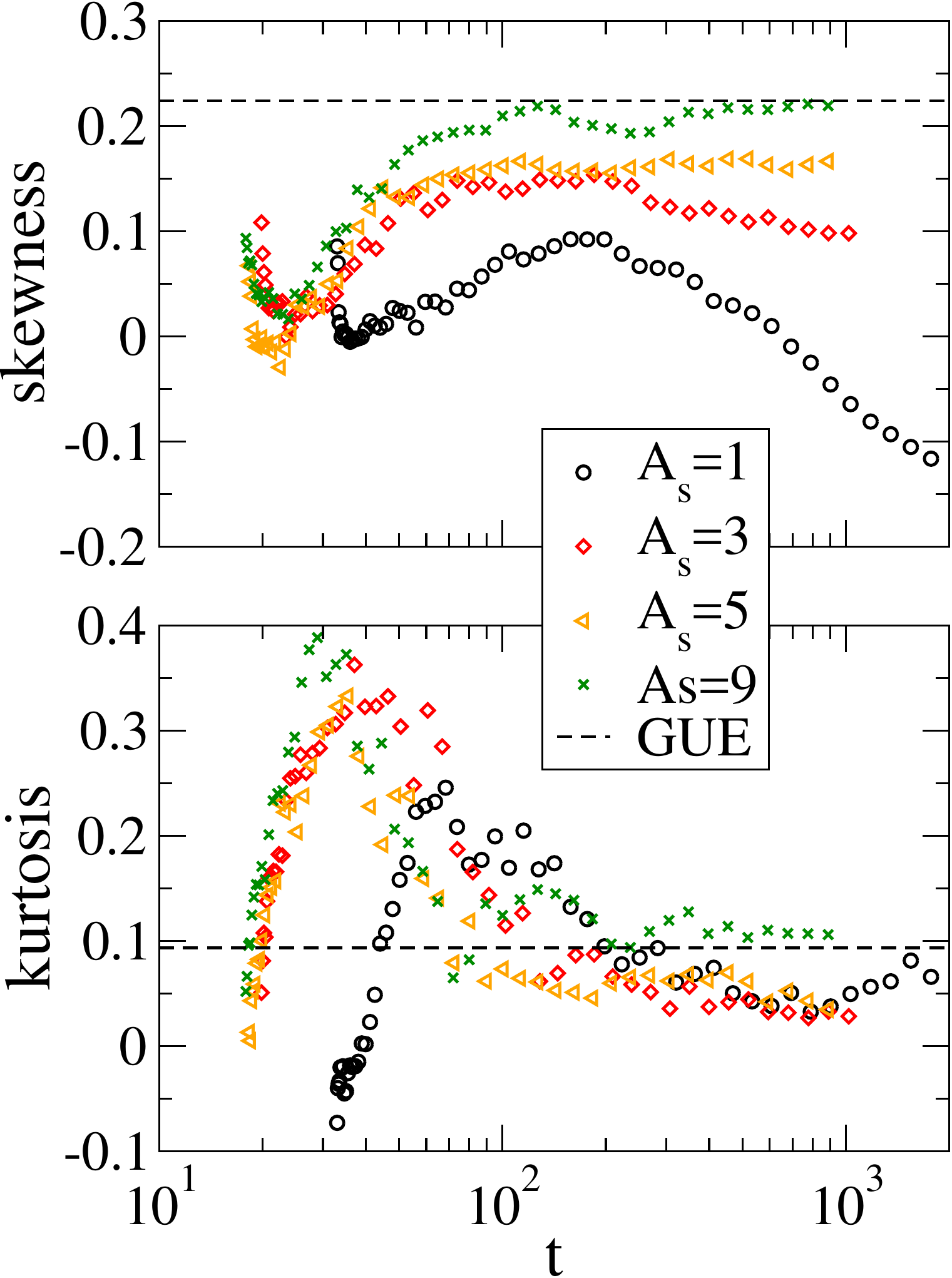}
	\caption{Evolution of the skewness and kurtosis for different shadowing
		strengths. Dashed lines are the values of the GUE-TW distribution expected for KPZ
		on a growing circular geometry~\cite{Takeuchi10}. Optimizations were not used
		in these simulations.}
	 \label{fig:skewkurt}
\end{figure}

If the KPZ ansatz given by Eq.~\eqref{eq:ansatz} holds, skewness and kurtosis
defined as 
\begin{equation}
\mathcal{S}=\frac{\av{r^3}_c}{\av{r^2}_c^{3/2}} \text{~~~~and~~~~}
\mathcal{K}=\frac{\av{r^4}_c}{\av{r^2}_c^2}, 
\end{equation}
respectively, should asymptotically provide universal cumulant ratios
$\mathcal{S}_\text{GUE}=0.2241$ and $\mathcal{K}_\text{GUE}=0.09345$
corresponding to the GUE-TW distribution~\cite{Prahofer00,Takeuchi2012}. From
Fig.~\ref{fig:skewkurt}, which shows skewness and  kurtosis against time for
different values of $A_s$, we observed a convergence to  the KPZ cumulants at
the longest times we investigated only for the weakest shadowing case.
Oppositely, convergence to a constant value was not observed for strong
shadowing effects, here  represented by $A_s=1$. In the moderate
	shadowing regime $A_s=3$ to $5$, the skewness exhibits a plateau during some time
	followed by a deviation at long times, as observed in the strong shadowing
	regime. So, our results for intermediate shadowing suggest that the KPZ regime
	could also be transient for $A_s=9$ or more generally for any finite $A_s$.
	However, this transient would be extremely large and the crossover to an
	unstable growth regime would not be attainable in doable simulations or experiments,
	in consonance with the experimental reports of Huergo \textit{et
		al}.~\cite{Huergo10,Huergo11,Huergo12}.

\begin{figure*}[ht]
	\centering
	\includegraphics[width=0.65\linewidth]{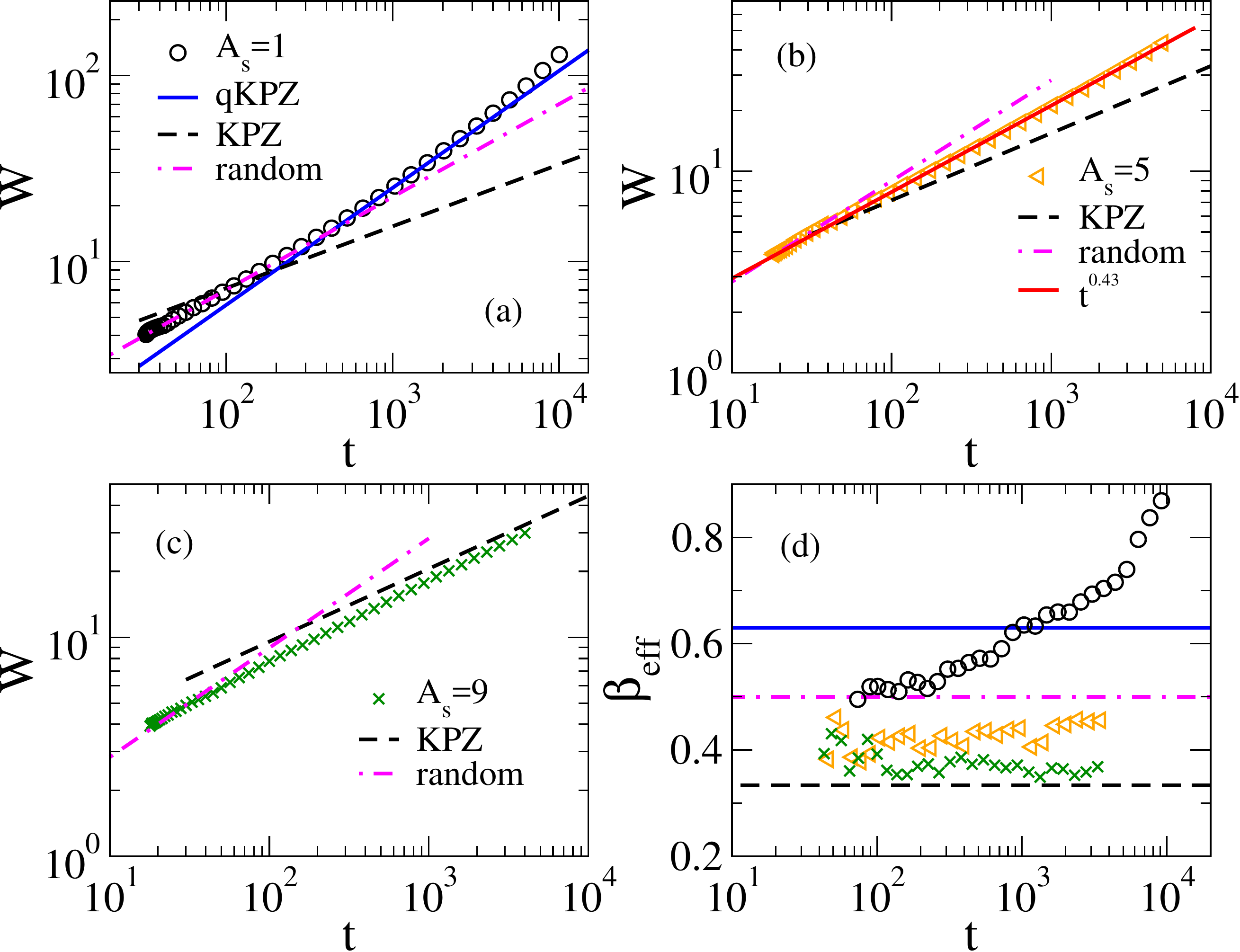}
	\caption{(a)-(c)Interface width against time for three levels of shadowing and
		the (d) effective growth exponent $\beta_\text{eff}=d\ln W/d\ln t$. Lines
		represent scaling laws of KPZ ($t^{1/3}$), qKPZ ($t^{0.633}$), and random
		deposition ($t^{1/2}$) growth regimes. Optimized simulations were used.}
	\label{fig:roughbetaeff_otim}
\end{figure*}

The interface width, Eq.~\eqref{eq:rugo}, is expected to scale as $W\sim
t^\beta$. Figures~\ref{fig:roughbetaeff_otim}(a)-(c) show the evolution of the
interface width for distinct shadowing strengths.  A local slope analysis
provides an effective growth exponent, defined as $\beta_\text{eff}=d\ln W/d\ln
t$~\cite{Alves13}, shown in Fig~\ref{fig:roughbetaeff_otim}(d). For weak
shadowing, represented by $A_s=9$ in Fig.~\ref{fig:roughbetaeff_otim}(c), we
obtain $\beta=0.34(1)$ at the late time plateau of the effective exponent, which
is in agreement with the KPZ value 1/3. For high shadowing, represented in
Fig.~\ref{fig:roughbetaeff_otim}(a), we see a initial regime consistent with
Poissonian random growth represented the exponent $\beta=1/2$~\cite{Barabasi95}.
For intermediate time a regime consistent with qKPZ exponent $\beta=0.633$,
manifested as a short plateau or, strictly, an inflection point in
the local exponent analysis, is observed. A clear deviation from the scaling
regime towards an unstable  growth, featured by a large growth exponent, is
observed at long times. This  transient qKPZ scaling is
	explained by metastable pinning points which appear in the surface due to
	the shadowing mechanism; see voids in Fig.~\ref{fig:agregados}(a). These pinning
	points play the role of a quenched disorder for some time. After 
	sufficient time, tips in the surface start to advance over the pinning points 
	and the dynamics undergoes a crossover to unstable growth. For moderate shadowing,
represented by $A_s=5$ in Fig.~\ref{fig:roughbetaeff_otim}(b), the growth
exponent quickly deviates from the KPZ and settles on a non-trivial exponent
$\beta\approx 0.43(2)$ for the whole time interval we investigated. We expect
that this would also deviate towards an unstable growth regime with large
$\beta$ if longer times could be attained.

\begin{figure}[ht]
	\centering
	\includegraphics[width=0.95\linewidth]{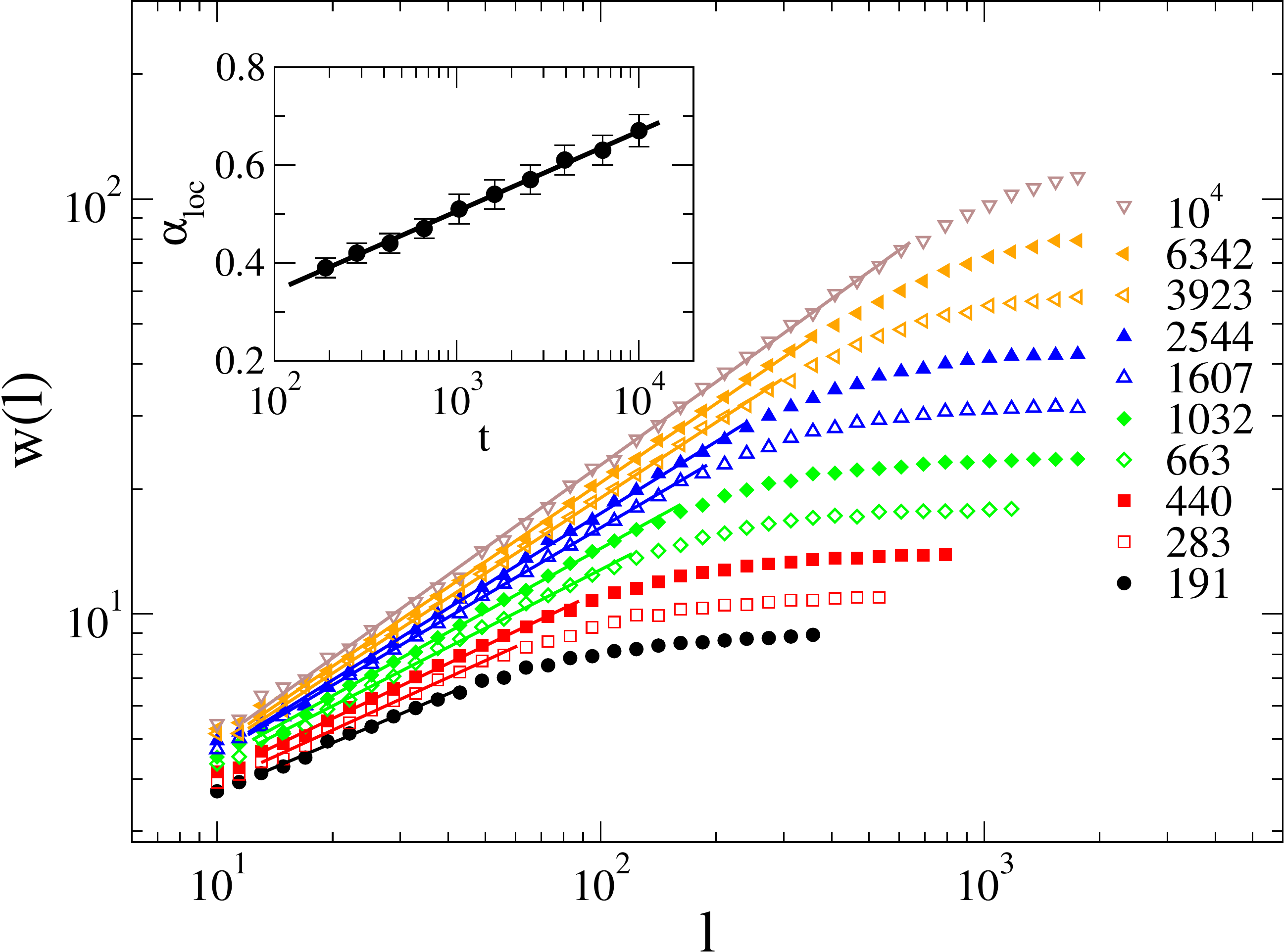}
	\caption{Local interface width for different times indicated in the legend using a high shadowing parameter $A_s=1$. Lines are power law regressions. Inset shows the local roughness (Hurst) exponent and the solid line is a logarithmic regression $\alpha_\text{loc} = 0.0158(9) + 0.071(1)\ln t$. Optimized simulations were used.}
	\label{fig:wtvslAs1}
\end{figure}

The large growth exponent can also indicate anomalous
scaling~\cite{Lopez1997a,Lopez,Castro1998,Nascimento} associated with a nonlocal
unstable growth~\cite{Nicoli2009a}. For example, the Mattew effect in which tips
grow faster than grooves amplifying differences in the interface, a mechanism
which is present in our model, can be associated with anomalous scaling and  be
misleadingly mixed up with qKPZ behavior~\cite{Nicoli2013}. In order to
investigate the presence of anomalous scaling, we analyzed the interface width
$w(l,t)$ within windows of size $l$, which is shown in Fig.~\ref{fig:wtvslAs1}
for the case of  high shadowing ($A_s=1$). For anomalous  scaling one expects
$w(l,t)\sim l^{\alpha_\text{loc}}t^\kappa$ where $\alpha_\text{loc}$ is the
Hurst exponent~\cite{Barabasi95} and $\kappa=( \beta z - \alpha_\text{loc} )/z$
where $\alpha_\text{loc}=\beta z$ restores the normal scaling. For short times
(and consequently short interface lengths) we see an apparent anomalous scaling
where the curves are shifted upwards as time is increased suggesting $\kappa>0$.
However, the local roughness exponents for different times,  shown in the inset
of Fig.~\ref{fig:wtvslAs1}, is increasing  logarithmically within the
range $0.39(2)<\alpha_\text{loc}<0.67(3)$ for the time interval $191<t<10^4$. Notice
that for the largest times we were able to simulate, the exponent is consistent
with the qKPZ exponent $\alpha_\text{qKPZ}=0.633$. The Hurst exponent continues
increasing for longer times and this system may present a super-roughness regime
with $\alpha_\text{loc}=1$ and $\alpha>1$ in the limit of very large times as
reported for other systems with morphological
instabilities~\cite{Lopez2005,Nicoli2013,Castro1998}.

\section{Concluding remarks}
\label{sec:conclu}

Recent experiments involving  growth of cell aggregates in dish
cultures~\cite{Huergo10,Huergo11,Huergo12} have provided the realization of the
exponents associated with the kinetic roughening of the  KPZ universality class
including the version with quenched noise (qKPZ) at the pinning
transition~\cite{Huergo14}. These findings are not trivial for two reasons.
Firstly, because cell growth is usually mediated by chemical signaling and
nutrient diffusion being, consequently, a non-local process at odds with the KPZ
equation and other models belonging to this class. Secondly, the realization of
qKPZ class without a tuning parameter to drive the systems towards a pinning
transition. This motivated us to investigate an off-lattice Eden model with
non-locality introduced by shadowing effects which mimic the competition for
nutrients diffusing from the outer region of the cell aggregate.

We observed a growth regime in agreement with the KPZ class, including the
exponents, dimensionless cumulant ratios, and corrections in the average
interface velocity, for weak shadowing effect which represents an
environment with many resources for cell replication. However, an initial random
growth that undergoes a crossover to unstable growth at asymptotic long times is
observed  for high shadowing effects (limited resources). Interestingly,  a
transient regime with exponents consistent with qKPZ class is observed in
limited resource regimes. Our results indicate that the KPZ regime is transient
and an unstable growth, possibly passing by a qKPZ regime, is expected at long
times. The qKPZ transient is due to the formation of metastable pinning points
caused by the shadowing mechanism introduced with the models. The actual
	universality class of this non-local Eden model needs further theoretical
	investigations. One natural possibility is to consider non-local KPZ
	equations~\cite{Mukherji1997,Katzav2002,Katzav2003}, as recently done
	in the context of non-universal biological growth~\cite{santalla17}.

Our numerical simulations support the experimental evidences for KPZ class in
cell aggregates grown in culture, in which a constant replacement of the culture
medium, providing thus sufficient nutrients and removing cell waste, was
necessary. The qKPZ exponents are expected to take place at the pining/depining
transition~\cite{Barabasi95} but they have  been reported in several
experiments, in which the pinning could be justified only
dynamically~\cite{Huergo14,Atis2013,Atis2015,Yunker2013}. The fact we observed a
transient behavior consistent with qKPZ in our non-local growth model without an
explicit quenched disorder can be an alternative or complementary explanation
for these experiments. Finally, our analysis was inspired in biological growth
but not restricted to it. So, we expect that our results can potentially be of
interest for other problems involving asymptotic unstable growth.

\begin{acknowledgments}
We thank Rodolfo Cuerno for useful comments and suggestions on the mansucript.
SNS acknowledges to Spanish Government for the financial support through
MINECO/AEI/FEDER (grant FIS2015-66020-C2-1-P) and MECD (``Jos\'e Castillejo''
program CAS15/00082). SCF acknowledges the Brazilian agencies CNPq and FAPEMIG
for the financial support.
\end{acknowledgments}


%

\end{document}